\documentclass[journal,comsoc,10pt]{IEEEtran}

\usepackage[T1]{fontenc}

\usepackage{cite}
\usepackage[colorlinks,
linkcolor=red,
anchorcolor=blue,
citecolor=green
]{hyperref}

\ifCLASSINFOpdf
\else
\fi
\usepackage{amsmath}
\usepackage{diagbox}

\usepackage{ amssymb }
\usepackage{amsfonts}
\usepackage{caption}
\usepackage{graphicx}
\usepackage{lettrine}
\usepackage{epstopdf}
\usepackage{textcomp,booktabs}
\usepackage[usenames,dvipsnames]{color}
\usepackage{colortbl}
\definecolor{mygray}{gray}{.9}
\definecolor{mypink}{rgb}{.99,.91,.95}
\definecolor{mycyan}{cmyk}{.3,0,0,0}
\usepackage{pifont}
\usepackage{subfigure}
\usepackage{multirow}
\usepackage{url}
\usepackage{color}
\usepackage{threeparttable}
\usepackage{setspace}
\usepackage{lineno}
\usepackage{tabularx}
\usepackage[dvipsnames,usenames]{color}

\usepackage{bm}

\usepackage{algorithm} 
\usepackage{algorithmic} 

\usepackage{dblfloatfix}

\usepackage[dvipsnames,usenames]{color}
\usepackage[usenames,dvipsnames]{color}

\definecolor{light-gray}{gray}{0.90}
\begin{document}
	\title{RIS-Enhanced Semantic Communications\\ Adaptive to User Requirements}

	\author{Peiwen Jiang, Chao-Kai Wen, Shi Jin, and Geoffrey Ye Li
			\thanks{P. Jiang and S. Jin are with the National
				Mobile Communications Research Laboratory, Southeast University, Nanjing
				210096, China (e-mail: PeiwenJiang@seu.edu.cn; jinshi@seu.edu.cn).}
			\thanks{C.-K. Wen is with the Institute of Communications Engineering, National
				Sun Yat-sen University, Kaohsiung 80424, Taiwan (e-mail: chaokai.wen@mail.nsysu.edu.tw).}
			\thanks{G. Y. Li is with the Department of Electrical and Electronic Engineering,
				Imperial College London, London SW7 2AZ, UK (e-mail: geoffrey.li@imperial.ac.uk).}}
	
	\maketitle
	\pagestyle{empty}  
	\thispagestyle{empty} 
%
%

\begin{abstract}

Semantic communication significantly reduces required bandwidth by understanding semantic meaning of the transmitted. However, current deep learning-based semantic communication methods rely on joint source-channel coding design and end-to-end training, which limits their adaptability to new physical channels and user requirements. Reconfigurable intelligent surfaces (RIS) offer a solution by customizing channels in different environments. 
In this study, we propose the RIS-SC framework, which allocates semantic contents with varying levels of RIS assistance to satisfy the changing user requirements. It takes into account user movement and line-of-sight obstructions, enabling the RIS resource to protect important semantics in challenging channel conditions. The simulation results indicate reasonable task performance, but some semantic parts that have no effect on task performances are abandoned under severe channel conditions. To address this issue, a reconstruction method is also introduced to improve visual acceptance by inferring those missing semantic parts. Furthermore, the framework can adjust RIS resources in friendly channel conditions to save and allocate them efficiently among multiple users.
Simulation results demonstrate the adaptability and efficiency of the RIS-SC framework across diverse channel conditions and user requirements.
	  
\end{abstract}

	\section{Introduction}


 \IEEEPARstart{S}{emantic} communication techniques have revolutionized information transmission by significantly saving transmission bandwidth and improving task performance \cite{bao2011towards}. Unlike traditional communication systems that aim to reduce the metric of bit-error rate, semantic systems transmit information tailored to meet users' specific requirements. This requires a content-aware transceiver that utilizes a precise knowledge base (KB) to interpret and optimize the transmission process. There have been some studies \cite{bao2011towards,gunduz2022beyond,shi2021semantic,qin2021semantic} to extract semantic features from the source contents and restore the essential elements, which mainly focus on preserving the intended meaning of the message by safeguarding its key semantic features even in the presence of partial corruption during transmission. Specifically, by employing deep learning (DL)-based joint source-channel coding architectures together with end-to-end training, the transmission process is optimized from the transmitter to the receiver, leading to accurate and efficient communication \cite{xie2020deep,9252948,wang2022wireless,liu2022task}.

However, the transition from bit-error reduction to improving semantic performance poses new challenges in system design. Additionally, physical channels are not always discernible due to the multipath effect and cannot be perfectly considered during the training process of semantic communication systems. To address these challenges, several modules have been revised to establish semantic transceivers \cite{jiang2022wireless}. For instance, in orthogonal frequency division multiplexing (OFDM) systems, explicit channel estimation \cite{yang2021deep} or redesigned channel estimation \cite{jiang2022adaptive} has been introduced to improve the accuracy of received semantic features. Modulation \cite{10012981,bo2022learning}, and peak-to-average power ratio \cite{10002903} have also been optimized to maximize semantic goals. In multiple-input multiple-output (MIMO) systems, the importance of semantic features can be leveraged to adaptively arrange parallel subchannels \cite{wu2022vision}. Furthermore, semantic resource allocation \cite{9763856, 10122232} has been proposed to enhance semantic performance in multiuser systems. Adaptive coding length has been explored to improve transmission efficiency under varying contents \cite{zhang2023semantic,9791409}. Additionally, reinforcement learning (RL) methods show promise in dealing with nondifferentiable physical channels \cite{beck2023model} and semantic similarity \cite{lu2021reinforcement}. Overall, existing works primarily focus on new architectures and training methods to enhance the flexibility of semantic communication systems.

Despite the advancements in semantic communication from a system perspective, challenges remain in harsh physical channels. Fortunately, reconfigurable intelligent surfaces (RIS) offer a promising solution for addressing the adaptability of semantic communication to dynamic channel environments \cite{9976945,9794416,9610122}. By controlling the RIS elements, the physical channel can be modified in real-time to meet semantic requirements by customizing the channel conditions for different semantic features. DL and RL techniques have been successfully applied to adjust RIS elements. For example, unsupervised DL methods have been proposed to design joint active and passive beamforming in RIS-assisted systems \cite{song2020unsupervised}. RL is commonly employed to optimize RIS elements for various goals, such as maximizing the sum-rate \cite{9794416} or jointly adjusting phase shift matrices with other modules \cite{gao2021machine,liu2020machine,huang2020reconfigurable}. RL has demonstrated its superiority in complex and unpredictable environments \cite{zhong2021ai}.

Inspired by RIS methods, we propose a novel communication transceiver called the RIS-enhanced semantic communication transceiver (RIS-SC). This transceiver is specifically designed to tackle transmission challenges arising from extreme channel conditions and varying user requirements. The RIS-SC transmitter integrates a semantic segmentation module and a joint source-channel encoder for image transmission. These components collaborate to transform a source image into a codeword carrying different semantic parts. Similar to 5G New Radio and WiFi systems, our proposed transceiver framework employs an OFDM module to send modulated symbols over the physical channels. At the receiver, a joint source-channel decoder aims to restore the original source image from the received symbols. What sets our approach apart is the introduction of an RL-based RIS-enhanced strategy to protect the transmit codeword from adverse channel effects. The core idea is to optimize the arrangement of different semantic parts on different subchannels and intelligently control the RISs based on changing user requirements, which act as rewards within the RL framework. By considering the interaction among the channel environment, the RIS controller, and user requirements, the RIS-SC system can dynamically safeguard important semantic features and achieve optimal task performance in varying channel conditions. For unimportant semantic parts, we propose a reconstruction method to enhance their visual representation. Although the reconstructed images may exhibit slight difference from the originals, they remain visually acceptable. In the RIS-SC framework, RIS resources are efficiently allocated in multiuser systems based on user requirements, satisfying their needs while maintaining the performance of other semantic parts.
  
The contributions of our study are as follows:

1) \textbf{High compatibility:} The weights in semantic transmission networks, such as semantic segmentation and encoder-decoder modules, do not require adjustment as long as the transmitted semantic parts are properly arranged and the channel conditions are customized for better task performance. This makes the proposed method easily combinable with different semantic transmission methods, as state-of-the-art studies commonly rely on the importance of different semantic parts and joint encoder-decoder schemes.

2) \textbf{Flexibility and low-cost:} Online adaptation of the proposed agents in the RL framework ensures the flexibility of the methods. With only a few training data collected online, the changing user requirements and varying channel conditions can be learned from feedback rewards. Once the user requirements are satisfied, the proposed method can save RIS resources for other users.

3) \textbf{Good vision:}  User requirements take top priority. Under poor channel conditions, unimportant semantic parts may be omitted and missing semantic parts can be inferred from available parts, resulting in visually reconstructed images. Since unimportant semantic parts are not a major concern for users, the reconstructed images are visually acceptable despite some difference from the originals.

	The rest of this paper is organized as follows. Section \uppercase\expandafter{\romannumeral2} introduces the system model, including conventional RIS-assisted MIMO systems and the classic RL-based RIS controller. The proposed transmission frameworks are presented in Section \uppercase\expandafter{\romannumeral3}, and their detailed architectures and training processes are discussed in Section \uppercase\expandafter{\romannumeral4}. Section \uppercase\expandafter{\romannumeral5} demonstrates the superiority of the proposed networks under varying channel conditions and changing user requirements, along with their ability to save limited resources and reconstruct poor semantic parts. Finally, Section VI concludes the paper.

\section{System Model}
\label{SystemModel}
In this section, the RIS-assisted transmission system is described first, including the channel model and the MIMO-OFDM transceiver. Then, a classic RL-based phase shift design is introduced. Finally, the existing challenges are listed to indicate the potential of a RIS-enhanced semantic communication framework.

\subsection{Channel Model for RIS-assisted Transmission}

This study considers a RIS-assisted single-cell MIMO-OFDM system, where the base station (BS) equipped with $N_{\rm B}$ antennas serves an user terminal (UT) with $N_{\rm U}$ antennas.  The RISs are uniform platform arrays (UPAs) of $M_{\rm r}$ rows,  $M_{\rm c}$ columns, therefore, there are $M=M_{\rm r} \times M_{\rm c}$ passive reflective elements in total.  We use $S$ RISs and the maximum delay spread of these channels is $L$-taps. The cascaded channel impulse response (CIR) at the $i$-th BS antenna and the $j$-th UT antenna can be expressed as
\begin{equation}
    \mathbf{h}_{i,j}=\mathbf{h}_{{\rm BU}, i,j}+\sum_{s=1}^S\sum_{m=1}^M \phi_{{\rm R_s}, m} \mathbf{h}_{{\rm BR_s}, i,j,m} \star \mathbf{h}_{{\rm R_sU}, i,j,m},
\end{equation}
where $\star$ represents the convolution operation,  $\mathbf{h}_{{\rm BU}, i,j} \in \mathbb{C}^{L \times 1}$ is the direct link from the BS to the UT, $\phi_{{\rm R_s}, m}$, $ \mathbf{h}_{{\rm BR_s},i,j,m}\in \mathbb{C}^{L \times 1}$, and  $\mathbf{h}_{{\rm RsU}, i,j,m}\in \mathbb{C}^{L \times 1}$ denote the reflection coefficient, BS-RIS and RIS-UT links of the  $m$-th element of the $s$-th RIS, respectively. For convenience, the amplitudes of $\phi_{{\rm Rs}, m}$ are set as one and their phase shifts  $\theta_{{\rm R_s}, m}$ are in $[0, 2\pi]$. The cascaded channel between different RISs is not considered because their powers are low. 

Due to the large number of RIS elements,  the elements in the same row are controlled together and the steering vector of the $j$-th row elements in the $i$-th RIS can be expressed as
\begin{equation}
    \Phi(\varphi_{R_{i,j}})=[e^{j\theta_1},\cdots,e^{j\theta_{M_c}}]=[1, e^{j \varphi},\cdots,e^{j(M_c-1) \varphi_i}],
\end{equation}
where $\varphi_i\in [0, \frac{1}{32}\pi,\cdots,\frac{63}{32}\pi]$ is adjusted by the BS. 

Assuming that the length of the cyclic prefix (CP) satisfies $L_{\rm CP} \geq L$,  received signal $\mathbf{Y}$ at the $k$-th subcarrier can be written in the frequency domain as
\begin{equation}
    \mathbf{Y}_k=\mathbf{H}_k\mathbf{V}_k\mathbf{X}_k+\mathbf{Z}_k,
\end{equation}
where 
\begin{equation}
   \mathbf{H}_k=\left[\begin{array}{ccc}
      {H}_{k,1,1}  &\cdots& {H}_{k,1,N_{\rm U}}  \\
       \vdots &\cdots& \vdots  \\
       {H}_{k,N_{\rm B},1} &\cdots& {H}_{k,N_{\rm U},N_{\rm B}} 
   \end{array}\right], 
\end{equation}
and ${H}_{k,i,j}$ is the channel frequency response (CFR) of the $k$-th subcarrier at  the $i$-th BS antenna and the $j$-th UT antenna, which is generated by $K$-point fast Fourier transform (FFT) of $\mathbf{h}_{i,j}$; $\mathbf{X}_k$ denotes the transmit OFDM symbols, $\mathbf{V}_k$ represents the precoding matrix based on singular value decomposition (SVD)  $\mathbf{H}_k=\mathbf{U}_k \mathbf{\Lambda}_k \mathbf{V}_k^H$, where $\mathbf{\Lambda}_k$ is a diagonal matrix, and $\mathbf{U}_k$ and $\mathbf{V}_k^H$ are unitary matrices. After applying the precoding matrix $\mathbf{V}_k$ and combining matrix $\mathbf{U}_k$, the estimated OFDM symbol can be written as 
\begin{equation}
\hat{\mathbf{X}}_k=\mathbf{\Lambda}_k \mathbf{X}_k+\mathbf{U}_k^H\mathbf{Z}_k.
\end{equation}
The above equation shows that the transmitted $\mathbf{X}_k$ is divided into multiple parallel subchannels, and the channel conditions are influenced by the diagonal elements in $\mathbf{\Lambda}_k$. These diagonal elements can be customized by utilizing RISs.

\subsection{Deep Reinforcement Learning for RIS}
The conventional algorithms for RIS rely on the accurate estimation of the cascaded channels and is with high-computation with the increasing of the RIS elements and antennas. Thus, deep learning is introduced to solve this issue in a model-free manner. Among these DL-based methods, deep RL is beneficial for the RIS control in a real-time environment, where the user movement and RIS elements changes the channel at the same time. In this section, the classic RL method is used for RIS control. 

\begin{figure}[h]
	\centering  

  {

  \includegraphics[width=0.8\linewidth]{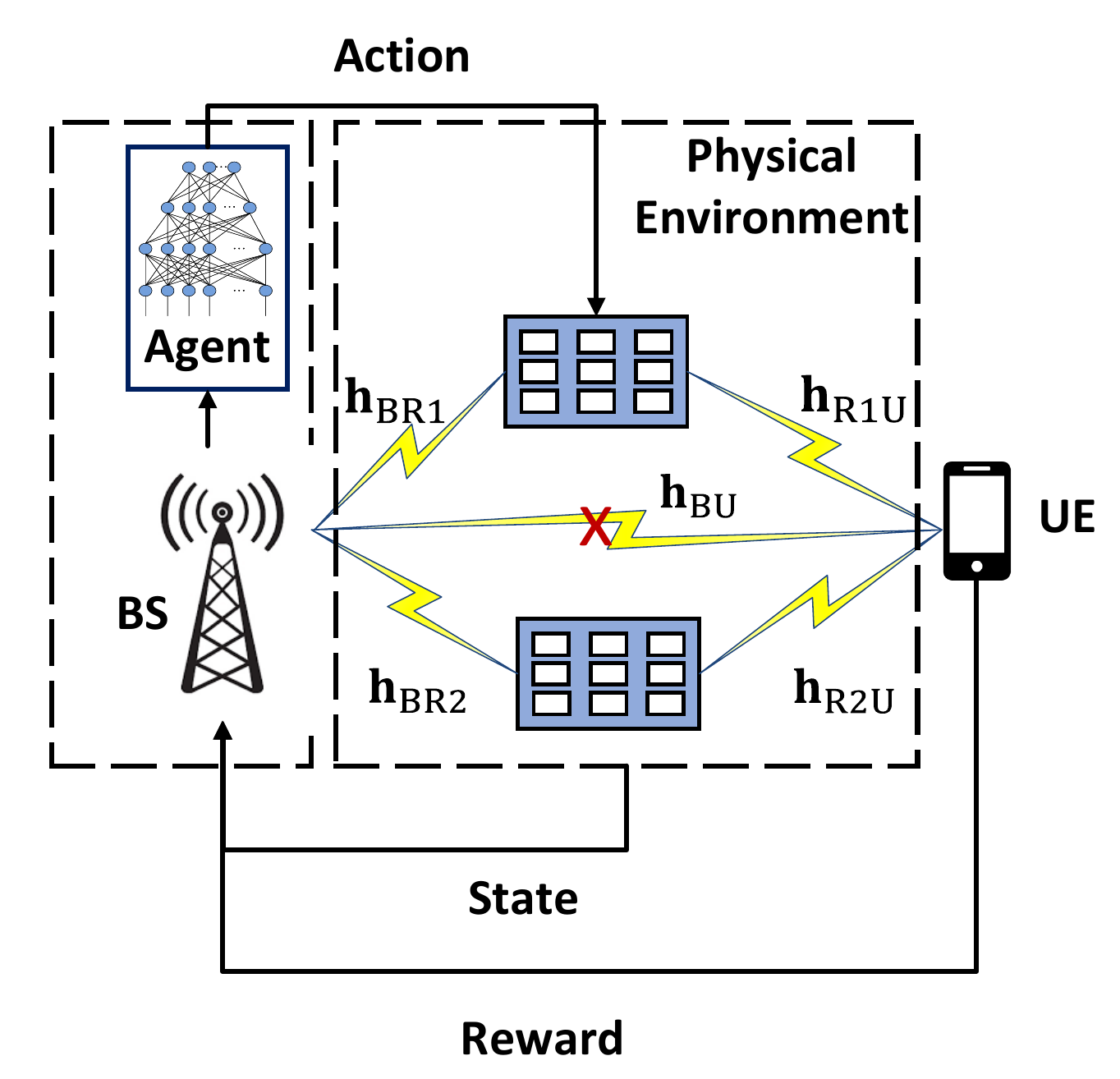}}
	\caption{ Framework of RL-based RIS control.}
	\label{DRL_RIS}
\end{figure}

As shown in Fig. \ref{DRL_RIS}, the RIS-assisted transmission system consists of the BS, the physical environment, and the UT. The RIS elements are controlled by the BS signal, allowing adjustment of the physical environment. The RL strategy employed in this system includes the following components:
\begin{itemize}
    \item State: The state represents the environmental characteristics observed by the agent at the BS. It includes the current channel estimation result and the phase shifts of the RIS elements.
    \item Action: The action is generated by the agents based on the current state. The action determines the changes to be made in the phase shifts of the RIS elements, thereby impacting the physical environment. In the subsequent time slots, the modified state is obtained by the BS, and a new action is sent.
    \item Reward: The transmission goals, such as maximum sum rate or SNR for different users, are calculated and serve as the rewards for the RL framework.
\end{itemize}

The RL-based RIS controller interacts with the physical environment and updates the agent according to the received reward.  In this study, a value-based RL method, specifically the deep Q-network (DQN), is employed. DQN selects actions from a discrete action space to maximize the reward and possesses a simple architecture. This method is chosen due to its low time consumption and fast convergence during online adaptation.

\subsection{Limitation of Existing Semantic Communications}
Semantic communication extracts the semantic features from the source and protect the important features according to the user requirements. The effectiveness of the semantic communication relies on the knowledge base (KB), which is implicitly contained in the trained parameters. Most studies establish the KB between the transmitter and receiver by end-to-end training, which is applicable only in channels with predictable characteristics. However, the fixed architecture and parameters of the existing semantic methods make them inflexible and less effective in extreme channel conditions.

RIS  offer a good compatibility with different semantic systems  as they are placed between the BS and the UT, allowing the reflected signals to enhance the transmission of semantic features. This implies that RIS can immediately improve the adaption of semantic communication by customizing the channel conditions. However, the challenges arise from the changing user requirements and the allocation of RIS resources, which pose difficulties for the RIS controller.

\section{Semantic segmentation-based image transmission framework}
\label{s3}

\begin{figure*}[h]
	\centering  

  {

  \includegraphics[width=0.9\linewidth]{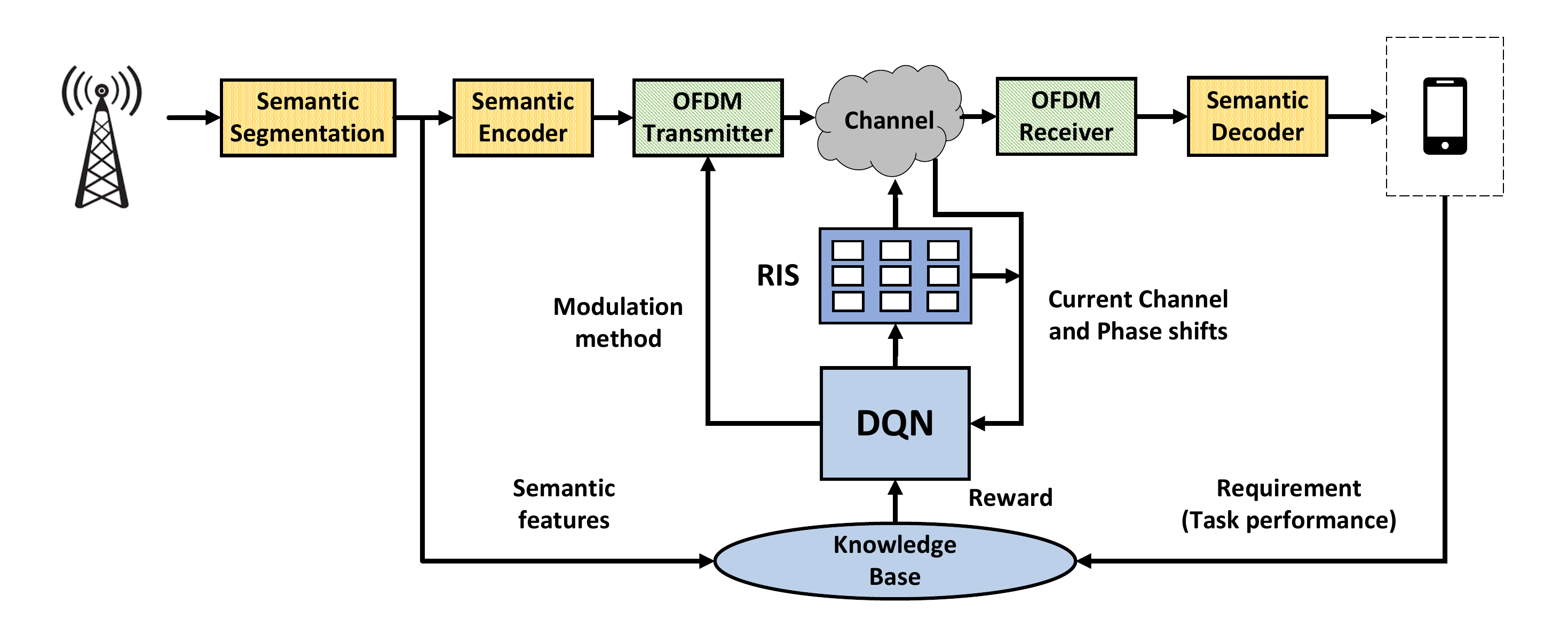}}
	\caption{ Overview of RIS-SC.}
	\label{Overall}
\end{figure*}
As illustrated in Fig. \ref{Overall}, the BS transmits an image to the UT. The proposed framework contains a semantic transmitter, a semantic receiver, and a RIS-assisted channel customization. The BS obtains the current estimated channel and utilizes this information to control the RIS elements, thereby modifying the channel based on the user's segmented semantic features. The DQN controller is guided by the reward from the KB. The KB possesses information about the user's requirements and the division of semantic features.  Concurrently, the DQN interacts with the current environment, encompassing channel conditions, RIS elements, and the modulation of semantic parts. In this section, we first introduce the framework of semantic image transmission. Subsequently, we describe the RL-based methods employed for RIS control. Finally, we discuss the image reconstruction method utilized to repair the faulty semantic parts.

\subsection{Semantic transceiver}
\label{s3a}

The source image at the $t$-th time slot $\mathbf{p}_t$ goes through a semantic segmentation, a semantic encoder, and an OFDM transmitter. The semantic segmentation $ {\tt Seg}(\cdot)$ divides the image into two semantic parts: background $\mathbf{p}_{\rm B}$ and object $\mathbf{p}_{\rm O}$, yielding
\begin{equation}
    [\mathbf{p}_{{\rm p},t}, \mathbf{p}_{{\rm O},t}]={\tt Seg}(\mathbf{p}_t),
\end{equation}
where ${\tt Seg}(\cdot)$ has a UNet-based semantic segmentation architecture \cite{ronneberger2015u}.

The semantic encoder-decoder shown in Fig. \ref{Overall} is based on SENet \cite{hu2018squeeze}, which adds an attention module in CNN blocks. The SENet based semantic encoder and decoder are denoted as ${\tt SC}_{\rm en}(\cdot)$ and ${\tt SC}_{\rm de}(\cdot)$, respectively. Transmitted bit sequence $\mathbf{b}_t$ can be expressed as  
\begin{equation}
\begin{aligned}
    \mathbf{b}_t&=[\mathbf{b}_{{\rm B},t}, \mathbf{b}_{{\rm O},t}]\\
    &=[{\tt Q}({\tt SC}_{\rm en}(\mathbf{p}_{{\rm B},t};\mathbf{W}_{{\rm en},1})),{\tt Q}({\tt SC}_{\rm en}(\mathbf{p}_{{\rm O},t};\mathbf{W}_{{\rm en},2}))],
    \end{aligned}
\end{equation}
where $\mathbf{W}_{{\rm en},1}$ and $\mathbf{W}_{{\rm en},2}$ contain trainable parameters of the encoders  of different semantic parts, and ${\tt Q}(\cdot)$ is a 8-bit quantization. After the transmission of the OFDM transceiver, the received bit sequence is $\hat{\mathbf{b}}$ and the image is restored as 
\begin{equation}
    \hat{\mathbf{p}}_t={\tt SC}_{\rm de}({\tt Q}^{-1}(\hat{\mathbf{b}}_t),\mathbf{W}_{\rm de}),
\end{equation}
where ${\tt Q}^{-1}(\cdot)$ is the dequantization operation and  $\mathbf{W}_{\rm de}$ is the set of trainable parameters of the semantic decoder.

Once trained, a semantic segmentation-based image transmission framework is established. However, the fixed network architecture lacks the ability to adapt to changing environments. For example, for object recognition, when the available bandwidth is limited, it becomes necessary to allocate more bandwidth to the transmission of the object part. Conversely, when the channel conditions are favorable, the background part should be restored effectively. It is important to note that user requirements are also subject to change, demanding a higher level of flexibility in semantic transmission.

In the proposed system, an OFDM system is employed to transmit bit sequence $\mathbf{b}_t$ at the $t$-th time slot. This bit sequence is encoded from semantic segmentation parts of the transmitted image.
The transmitted bit sequence is then modulated into transmission OFDM symbol  $\mathbf{X}_{t}$. Similar to Eq. (5), the received OFDM symbol at the $k$-th subcarrier can be expressed as
\begin{equation}
    \hat{\mathbf{X}}_{k,t}=\mathbf{\Lambda}_{k,t} \mathbf{X}_{k,t}+\mathbf{U}_{k,t}^H\mathbf{Z}_{k,t},
\end{equation}
where the SVD operation is applied as  $\mathbf{H}_{k,t}=\mathbf{U}_{k,t} \mathbf{\Lambda}_{k,t} \mathbf{V}_{k,t}^H$ and the current channel $\mathbf{H}_t=[\mathbf{H}_{1,t},\cdots,\mathbf{H}_{k,t}]$ at the $t$-th time slot is affected by the RIS steering vectors, $\Phi_{{\rm R},t}=[\Phi_{{\rm R}_{1,1},t},\cdots,\Phi_{{\rm R}_{S,M_{\rm r}},t}]$. After passing through the OFDM transceiver and the current channel, the estimated OFDM symbol $\hat{\mathbf{X}}_t$ is converted to estimated bit sequence $\hat{\mathbf{b}}$. Finally, the received image is reconstructed.
Consequently, the training data can be collected along with the time. 

In the subsequent sections, we explore RIS-enhanced methods that address the challenges posed by changing channels and varying user requirements. These methods hold promise as they offer a potential solution without the need to update trainable parameters or revise network architectures in a semantic transmission system.

\subsection{RL strategies for RIS-SC}

 The agent receives inputs of current channel $\mathbf{H}_t$ and the steering vectors of all the RISs, $\Phi_{{\rm R},t}$.
 It then outputs an action $\Delta \varphi_{{\rm R}_{i, j},t+1}$, which is selected from a set of five actions, $[-\frac{3}{32}\pi, -\frac{1}{32}\pi,0,\frac{1}{32}\pi,\frac{3}{32}\pi]$.
 Thus, the steering vector of RIS  at the next time slot is adjusted as $\Phi_{{\rm R}_{i,j}}(\varphi_{{\rm R}_{i,j},t}+\Delta\varphi_{{\rm R}_{i,j},t+1})$.  The entire process can be expressed as 
 \begin{equation}
     \Delta\varphi_{{\rm R}_{i,j},t+1}=f_{{\rm R}_{i,j}}(\mathbf{H}_t,\Phi_{{\rm R},t}; \mathbf{W}_{{\rm R}_{i,j}}), \label{eqA}
 \end{equation}
where $f_{{\rm R}_{i,j}}(\cdot,\mathbf{W}_{{\rm R}_{i,j}})$ represents the agent responsible for controlling the phase shift of the $j$-th row phase shift of the  $i$-th RIS. The trainable parameters of this agent are denoted by $\mathbf{W}_{{\rm R}_{i,j}}$. 

In addition to controlling the RIS elements, an agent is utilized to determine the arrangement of transmission bits. For instance, in a $2\times 2$ MIMO system with a maximum of two streams, the bits carrying different semantic parts, $\mathbf{s}_{\rm B}$ and $\mathbf{s}_{\rm O}$, can be 4-quadrature amplitude modulation (QAM) modulated and transmitted on different streams. Alternatively, these bits can be 16-QAM modulated and transmitted on one stream while the other stream is blocked. Therefore,  the agent for the $i$-th semantic part needs to determine which stream is chosen for transmission, denoted as $a_{{\rm Mod},i,t+1}$.  Along with the current channel and RIS steering vectors, the current modulation scheme $\mathbf{a}_{{\rm Mod},t}=[a_{{\rm Mod,1},t},a_{{\rm Mod},2,t}]$ is also inputted into the agent as a reference.  The whole process can be expressed as 
\begin{equation}
a_{{\rm Mod},i,t+1}=f_{{\rm Mod}, i}(\mathbf{H}_t,\Phi_{\rm R,t},\mathbf{a}_{{\rm Mod},t}; \mathbf{W}_{{\rm Mod},i}), \label{eqB}
\end{equation}
where $f_{\rm Mod, i} (\cdot;\mathbf{W}_{\rm Mod,i} )$ represents the agent for the i-th semantic part, which utilizes trainable parameters $\mathbf{W}_{\rm Mod,i}$. The output, $\mathbf{a}_{{\rm Mod},t}$, determines the selection of the transmission stream. Once the number of bits in the stream is determined, the appropriate modulation method is used to convey these bits across the subcarriers.

\begin{figure*}[t]
	\centering  

		{
	
		{\includegraphics[width=0.9\linewidth]{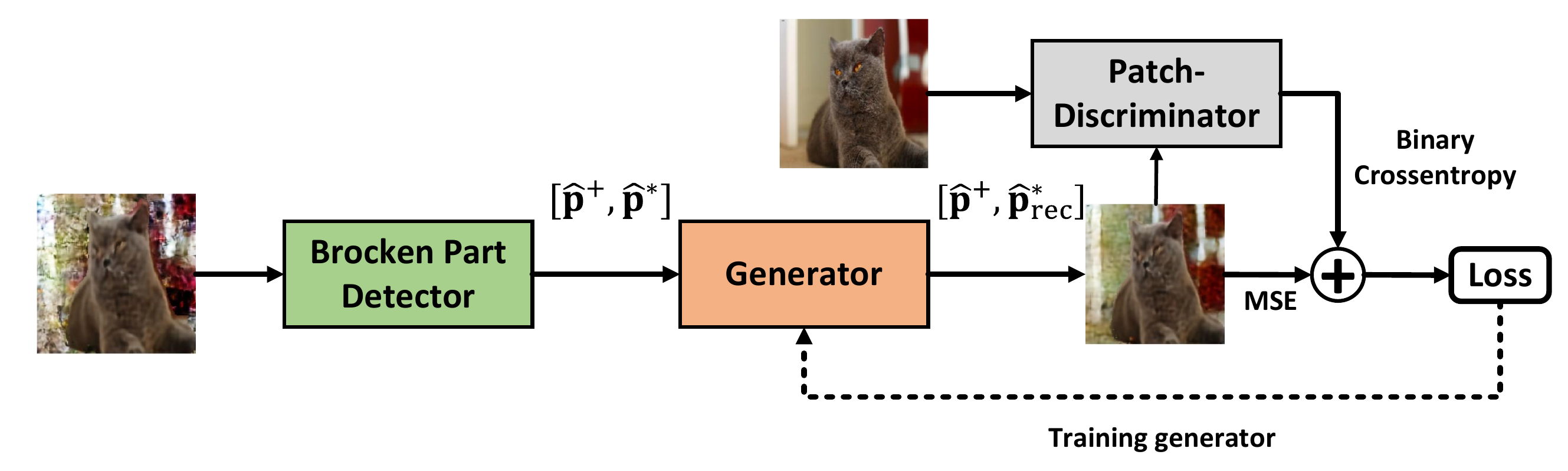}}}
 
	\caption{Framework of the broken part detector and the generator for image reconstruction.}
	\label{Rec_fra}
\end{figure*}
In summary, through the interaction among the channel environment, the agents, and the user requirements, the proposed RIS-SC framework demonstrates flexibility and the ability to handle extreme environments without the need for retraining the semantic transceiver. Such retraining would consume significant computational resources and is impractical for online adaptation.

\subsection{Allocation for RIS resources}

The method mentioned above utilizes all the resources of the RIS for semantic transmission. However, the resources of the RIS are limited within a specific area and are shared among different users. This implies that the RISs can be reserved for other applications once the user's requirements are fulfilled. Two conditions are considered in this scenario:

 \begin{itemize}
     \item Users are served by different BSs. Therefore, the RIS resources cannot be allocated jointly.

     \item Users are simultaneously served by the same BS, and the allocation of RIS resources among these users is managed by the BS.
 \end{itemize}

 Since the BS does not have knowledge about the requirements and channel conditions of users served by other BSs in the first condition, it only obtains the current usage of the RIS elements and aims to conserve the RIS resources in the subsequent action. The $j$-th row of the $i$-th RIS is chosen at the next time slot when the output $a_{{\rm C},i,j,t+1}$ is 1; otherwise,  $a_{{\rm C},i,j,t+1}$ is 0. The output can be determined using 
\begin{equation}
\begin{aligned}
   \mathbf{a}_{{\rm C},i, t}= [a_{{\rm C},i,1,t+1},\dots,a_{{\rm C},i,M_r,t+1}]\\=f_{{\rm C}, i}(\mathbf{H}_t,\Phi_{{\rm R},t},\mathbf{a}_{{\rm C}, t}; \mathbf{W}_{{\rm C},i}),
   \end{aligned}
\end{equation}
where $\mathbf{a}_{{\rm C}, t}=[a_{{\rm C},1,t},\dots,a_{{\rm C},M_c,t}]$ represents the current chosen rows for all RISs and $\mathbf{W}_{\rm C,i}$ includes the trainable parameters of agent $f_{{\rm C}, i}(\cdot)$. To save the RIS resources,  the total number of utilized RIS rows is incorporated into the reward function as a penalty term, represented by $-5\Sigma_{i=1}^{M_r} \Sigma_{i=1}^{M_c} a_{{\rm C},i,j,t}$.

In the second scenario, where the requirements and channel conditions of users served by the same BS are known, all the agents for these users can be trained together. For instance, in a 4×4 multiuser MIMO system with the BS having four antennas and two users equipped with two antennas each, the two users transmit different pictures, $\mathbf{p}_{1,t}$ and $\mathbf{p}_{2,t}$, simultaneously. The training data is collected according to their respective requirements. The other agents for RIS phase shifts and modulation methods gather and undergo similar training processes. Generally, the task performance of the two users guides all the agents.

Through these RIS allocation strategies, the RIS-SC framework can effectively balance the task performance while operating within the limitations of available resources. The important semantic parts are transmitted with priority and some unimportant parts can be omitted to save transmission resources for other users. 

\subsection{Reconstruction for Blocked Semantic Parts}
To address the problem of unimportant semantic part that are not well reconstructed at the receiver due to changing channels and limited transmission resources, a solution is proposed involving the generation of the broken semantic parts based on the well-transmitted parts. In Fig. \ref{Rec_fra}, the reconstruction process consists of two steps: broken part detection and reconstruction.

The broken part detection involves  a similar problem as the semantic segmentation and thus the segmentation network ${\tt Seg}(\cdot)$ is employed to divide the estimated picture into two parts, good part $\hat{\mathbf{p}}^+$ and bad part $\hat{\mathbf{p}}^*$. This can be expressed as
\begin{equation}
    [\hat{\mathbf{p}}^+,\hat{\mathbf{p}}^*]={\tt Seg}(\hat{\mathbf{p}}),
\end{equation}
where $\hat{\mathbf{p}}$ is the estimated picture. Then, the bad part $\hat{\mathbf{p}}^*$ is removed from the received image and reconstructed using a generator network $G(\cdot)$. The process can be represented as
\begin{equation}
    \hat{\mathbf{p}}_{\rm rec}=[\hat{\mathbf{p}}^+,\hat{\mathbf{p}}^*_{\rm rec}]=G([\hat{\mathbf{p}}^+,\hat{\mathbf{p}}^*]),
\end{equation}
where reconstructed part $\hat{\mathbf{p}}^*_{\rm rec}$ is combined with the good part and the whole image has a better
visual result. In extreme condition, bad part $\hat{\mathbf{p}}^*$ is covered by the noise. Reconstructed
$\hat{\mathbf{p}}^*_{\rm rec}$ is totally guessed by the generator but it makes the image acceptable to the user compared
to an image with a high-noised part.

\section{Detailed architecture and training process }
In this section, the modules in RIS-SC are introduced in detail. Then, the offline and online training process of these modules are described.
\subsection{Detailed network architectures}
\begin{figure*}[t]
	\centering  

		\subfigure[]{
		\label{DQN_phase}
		{\includegraphics[width=0.9\linewidth]{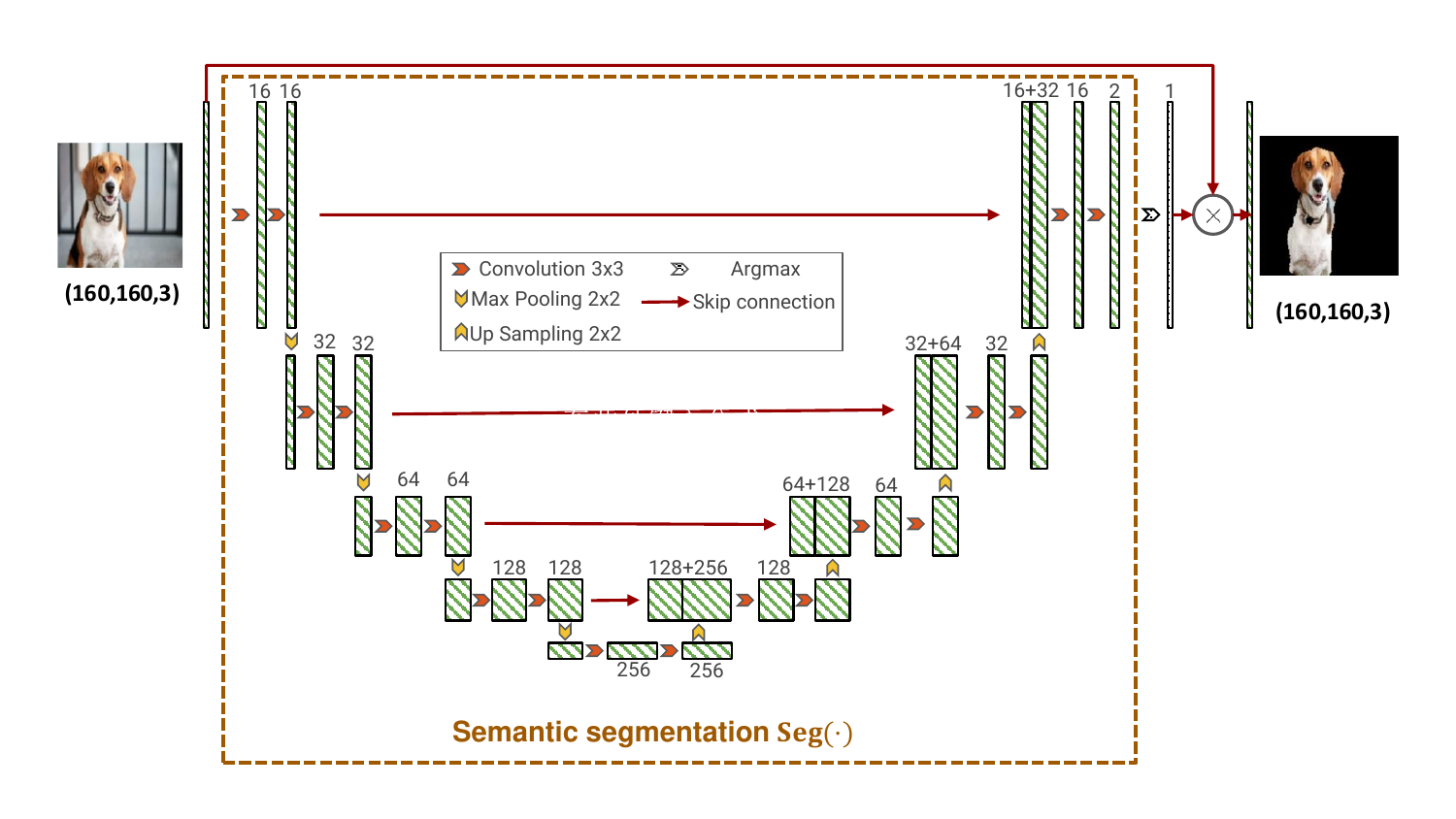}}}
  \subfigure[]{
		\label{DQN_CQI}
		{\includegraphics[width=0.9\linewidth]{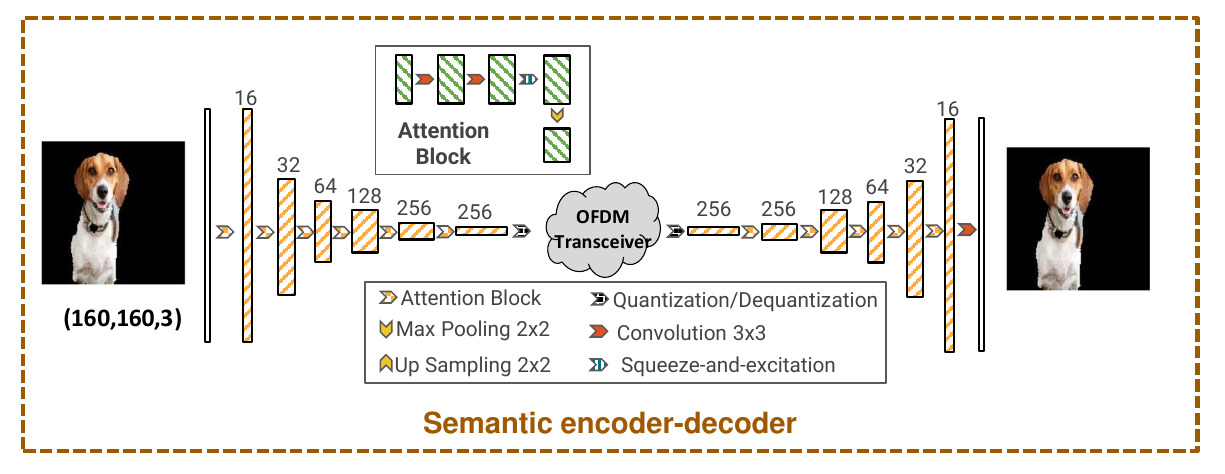}}}

	\caption{Architectures of the semantic segmentation and encoder-decoder. The number of output channels  is marked above each CNN-based operation. }
	\label{Detailed modules}
\end{figure*}

\textbf{Semantic segmentation network ${\tt Seg}(\cdot)$:} As shown in Fig. \ref{Detailed modules}(a), this architecture is consists of 3×3 convolution CNN operations with a ReLU activation function, as well as a down/upsampling operation.  The number of channels  is marked above each CNN operation. Finally, the output of the last CNN operation is converted into (160, 160, 2) using a softmax activation function,  enabling the classification of each pixel into background and object categories. The UNet-based skip connections between CNN blocks are are employed between the CNN blocks to learn the semantic features in different aspects. Consequently,   ${\tt Seg}(\cdot)$ can segment the input image into two semantic parts. In the example shown in Fig. \ref{Detailed modules}(a),  the input image has a dimension of (160, 160, 3) and its object part is needed. The argmax function is applied to the output of ${\tt Seg}(\cdot)$ to generate  (160, 160, 1), where the location of the object pixel is marked as 1 and the others are marked as 0. This binary matrix is used to mask the original image, preserving only the object part..

\textbf{Semantic encoder and decoder ${\tt SC}_{\rm en}(\cdot)$,  ${\tt SC}_{\rm de}(\cdot)$:} As shown in Fig. \ref{Detailed modules}(b), the encoder comprises six attention blocks, while the decoder consists of five attention blocks. Each  attention block consists of two $3 \times 3$ convolution operations with Relu activation function, a squeeze-and-excitation operation, and a two times down/upsampling operation. The last convolution operation in the decoder outputs the decoded image.

\begin{figure*}[t]
	\centering  

		{
		\label{DQN_eff}
		{\includegraphics[width=1\linewidth]{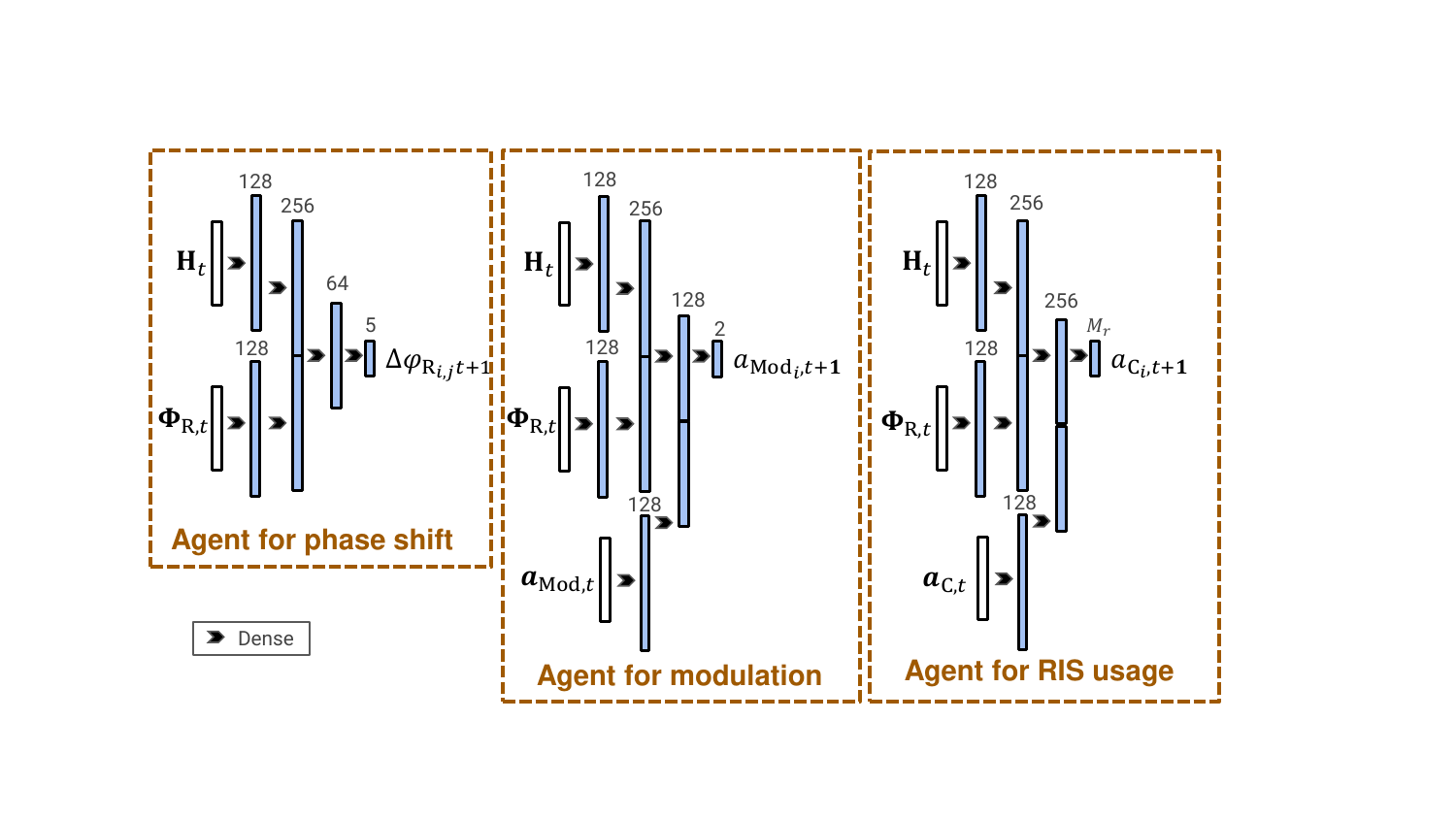}}}
 
	\caption{Architectures of agents for phase shifts, modulation, and RIS usage, respectively. The number of output neurons is marked above each dense layer.}
	\label{Agents}
\end{figure*}

\textbf{Agents for RIS phase shift $f_{{\rm R}_{i,j}}(\cdot)$:} As shown the left of Fig. \ref{Agents}, the current information, including $\mathbf{H}_t$, $\Phi_{\rm R,t}$, is fed into dense layers with 128 neurons.  The outputs are then combined and inputted into a dense layer with 256, 64, and 5 neurons. All hidden layers use Relu activation function, and the final layer outputs five values with a softmax activation function. 

\textbf{The agents for modulation $f_{{\rm Mod},{i}}(\cdot)$:} As shown in the middle of Fig. \ref{Agents}, apart from  $\mathbf{H}_t$, $\Phi_{\rm R,t}$, the  previous modulation method serves as a reference input into a dense layer with 128 neurons.  The output is  the second-to-last layer, which is further connected to a dense layer with 128 and 2 neurons.  All the hidden layers use Relu activation function.  In a two-stream scenario, the last layer outputs two values using a softmax activation function, but it can be expanded to accommodate more streams.

\textbf{The agents for RIS usage  $f_{{\rm C},{i}}(\cdot)$:} The agent to manage the RIS usage is  shown in Fig \ref{Agents}.  The architecture is similar to the agent $f_{\rm Mod,i}(\cdot)$, while  the last layer uses sigmoid activation function. The number of output values corresponds to the number of rows in the RIS. A hard decision is made to convert the output into either 0 or 1. A value of 0 indicates that the corresponding row is reserved for other users while a value of 1 signifies that the row is selected and adjusted in the subsequent time slot.

\textbf{Broken part detector ${\tt Seg}(\cdot)$ and generator $G(\cdot)$:} The broken part detector employs the same structure as the semantic segmentation network, as it also classifies the pixels into two categories: good and bad.  The generator, on the other hand, is a standard UNet model, with the last CNN block producing a dimension of (160, 160, 3) to restore the image.

\subsection{Training process}

\textbf{Semantic transceiver:}
To automatically extract semantic features,  the semantic encoder-decoder is trained  in an end-to-end manner. For fast converge, 8-bit quantization is not applied initially, and the transmitted real numbers are passed through an additive white Gaussian noise channel $h(\cdot)$  with an SNR of 10 dB. The loss function is mean-squared error (MSE) with learning rate of 0.001. The training process is as follows
\begin{equation}
\begin{aligned}
    &(\hat{\mathbf{W}}_{{\rm en},1},\hat{\mathbf{W}}_{{\rm en},2},\hat{\mathbf{W}}_{{\rm de}})=\\ &\mathop{\arg\min}\limits_{\mathbf{W}_{{\rm en},1},\mathbf{W}_{{\rm en},2},\mathbf{W}_{{\rm de}}} ||\mathbf{p}_t-SC_{\rm de}(h(SC_{\rm en}(\mathbf{p}_t)))||^2. 
    \end{aligned}
\end{equation}
Once the network is trained, it undergoes finetuning with 8-bit quantization when the channel $h(\cdot)$ has a bit-error rate of 0.01. The learning rate for finetuning is set to 0.0001 and the process is  
\begin{equation}
\begin{aligned}
    &(\hat{\mathbf{W}}_{{\rm en},1},\hat{\mathbf{W}}_{{\rm en},2},\hat{\mathbf{W}}_{{\rm de}})= \\
    &\mathop{\arg\min}\limits_{\mathbf{W}_{{\rm en},1},\mathbf{W}_{{\rm en},2},\mathbf{W}_{{\rm de}}} ||\mathbf{p}_t-SC_{\rm de}(Q^{-1}(h(Q(SC_{\rm en}(\mathbf{p}_t)))))||^2. 
    \end{aligned}
\end{equation}

\textbf{Agents:}
Two classic requirements are considered for designing the reward:

1) In the first reward, accurate classification of transmitted objects in the image is crucial. To assess the transmission quality of objects, a VGG-based classifier \cite{simonyan2014very} is employed.  The accuracy of the classifier on the received image is $f_{\rm ACC}(\hat{\mathbf{p}}_t)$, or simply ACC in convenience when testing. Then, the global quality of the image is measured using  MSE, represented by $f_{\rm MSE}(\mathbf{p}_t, \hat{\mathbf{p}}_t)$. This reward is given by
\begin{equation}
\begin{aligned}
    &R_{\rm ACC}(\hat{\mathbf{p}}_t,\mathbf{p}_t)=\\&\left\{ \begin{aligned}
        &10f_{\rm ACC}(\hat{\mathbf{p}}_t)-10 \log_{10}(f_{\rm MSE}(\mathbf{p}_t, \hat{\mathbf{p}}_t)),&f_{\rm ACC}(\hat{\mathbf{p}})>0.85,\\
       &10f_{\rm ACC}(\hat{\mathbf{p}}_t)-100, & f_{\rm ACC}(\hat{\mathbf{p}})\leq 0.85,
    \end{aligned}\right.
    \end{aligned}
\end{equation}
where $ 10f_{\rm ACC}(\hat{\mathbf{p}})-100$ represents a penalty when the object classifier accuracy threshold is set to 0.85 while the optimal performance is $f_{\rm ACC}(\hat{\mathbf{p}})=f_{\rm ACC}({\mathbf{p}})=0.95$.

2) In the second reward, there is no requirement for object classification and the focus is solely on the MSE performance
\begin{equation}
    R_{\rm MSE}(\hat{\mathbf{p}}_t,\mathbf{p}_t)=-10 \log_{10}(f_{\rm MSE}(\mathbf{p}_t,\hat{\mathbf{p}}_t)).
\end{equation}
The RIS-SC system that prioritizes object classification based on $R_{\rm ACC}$, is referred to as RL-ACC while the system that emphasizes global image quality based on $R_{\rm MSE}$ is called RL-MSE. In cases where the user requirement is unknown, the conventional sum rate is used as the reward and the RIS-SC depends on this reward is referred to as RL-Rate.

The training process for these agents is divided into two phases: 

1) Initially, all the transmission resources are used for training. The UT receiver, which is aware of the transmitted images, calculates $f_{\rm ACC}(\hat{\mathbf{p}}_t)$, $f_{\rm MSE}(\mathbf{p}_t, \hat{\mathbf{p}}_t)$, or sum rate. For example, the agent responsible for  the $j$-th row of the $i$-th RIS is rained when the user requirement is object recognition. The training data includes $[R_{\rm ACC}(\hat{\mathbf{p}}_t,\mathbf{p}_t),$ $\mathbf{H}_t,$ $\Phi_{\rm R,t},$ $, \Delta\varphi_{{\rm R}_{i,j},t+1}]$. The batch size is eight, and a total of  1280 images are transmitted for training. After 100 training epochs, these agents demonstrate good performance and can be deployed.

2) When transmission begins, the real image $\mathbf{p}_t$ is unknown, and the reward cannot be calculated. Therefore,  the BS sends a known image per time interval for training. Once the known image is received, the new reward can be calculated and the agents are finetuned to adapt to changing channel environment and user requirements.

For multiuser systems where multiple users are served by the same BS, their requirements can be considered together. The training data includes
$[R_{\rm ACC}(\hat{\mathbf{p}}_{1,t},\mathbf{p}_{1,t})+R_{\rm ACC}(\hat{\mathbf{p}}_{2,t},\mathbf{p}_{2,t}),$
$\mathbf{H}_t,\Phi_{\rm R,t}, $ 
$\Delta\varphi_{{\rm R}_{i,j},t+1}]$ when the first user transmits $\mathbf{p}_{1,t}$, the second user transmits $\mathbf{p}_{2,t}$, both requiring object recognition.

\textbf{Image reconstruction:}
The training data is collected during the training of the aforementioned agents. The bad part refers to the semantic part transmitted under poor subchannels, with an MSE larger than 0.05. Conversely, the good part represents the semantic part transmitted under favorable subchannels. The pixels corresponding to the bad part are marked, and a segmentation network is trained to classify all pixels into good and bad parts using binary cross-entropy as the loss function.

 The generator is trained with a patch discriminator $D(\cdot)$\cite{wang2019few} and MSE. The overall loss function is defined as
\begin{equation}
    L_{\rm G}= L_{\rm BC}(\mathbf{1}, D(\hat{\mathbf{p}}_{\rm rec}))+100*||\hat{\mathbf{p}}_{\rm rec}-\mathbf{p}||^2,
\end{equation}
where  $L_{\rm BC}$ is the binary crossentropy loss function. Following a step of generator training,  the discriminator is updated with the loss function as
\begin{equation}
    L_{\rm D}=L_{\rm BC}(\mathbf{0},D(\hat{\mathbf{p}}_{\rm rec}))+L_{\rm BC}(\mathbf{1},D(\mathbf{p})).
\end{equation}

Based on the module descriptions and training processes provided, the proposed framework's components are outlined. The semantic transceiver is trained and fixed without insights into varying channel conditions and user requirements. Consequently, computational resources are conserved, as retraining large semantic networks can be resource-intensive. Online adaptation occurs through the interaction between the agents and the environment. With controllable RIS elements, the performance of semantic transmission can be enhanced in extreme channel environments, where offline learning alone may not yield optimal performance.

\section{Numerical Results}
\label{s5}
After introducing the simulation settings, we compare the proposed SC-RIS method with competing methods and highlight the benefits of incorporating RIS. Furthermore, we analyze the allocation of RIS resources and demonstrate through transmission examples how user requirements can guide the agent to save resources without compromising task performance. We also test the adaptability of RL-based methods to changing user requirements. Finally, we discuss the reconstruction of images when certain semantic parts are blocked.

\subsection{Simulation Settings}

\begin{figure}[h]
	\centering  
    \subfigure[]{\includegraphics[width=0.9\linewidth]{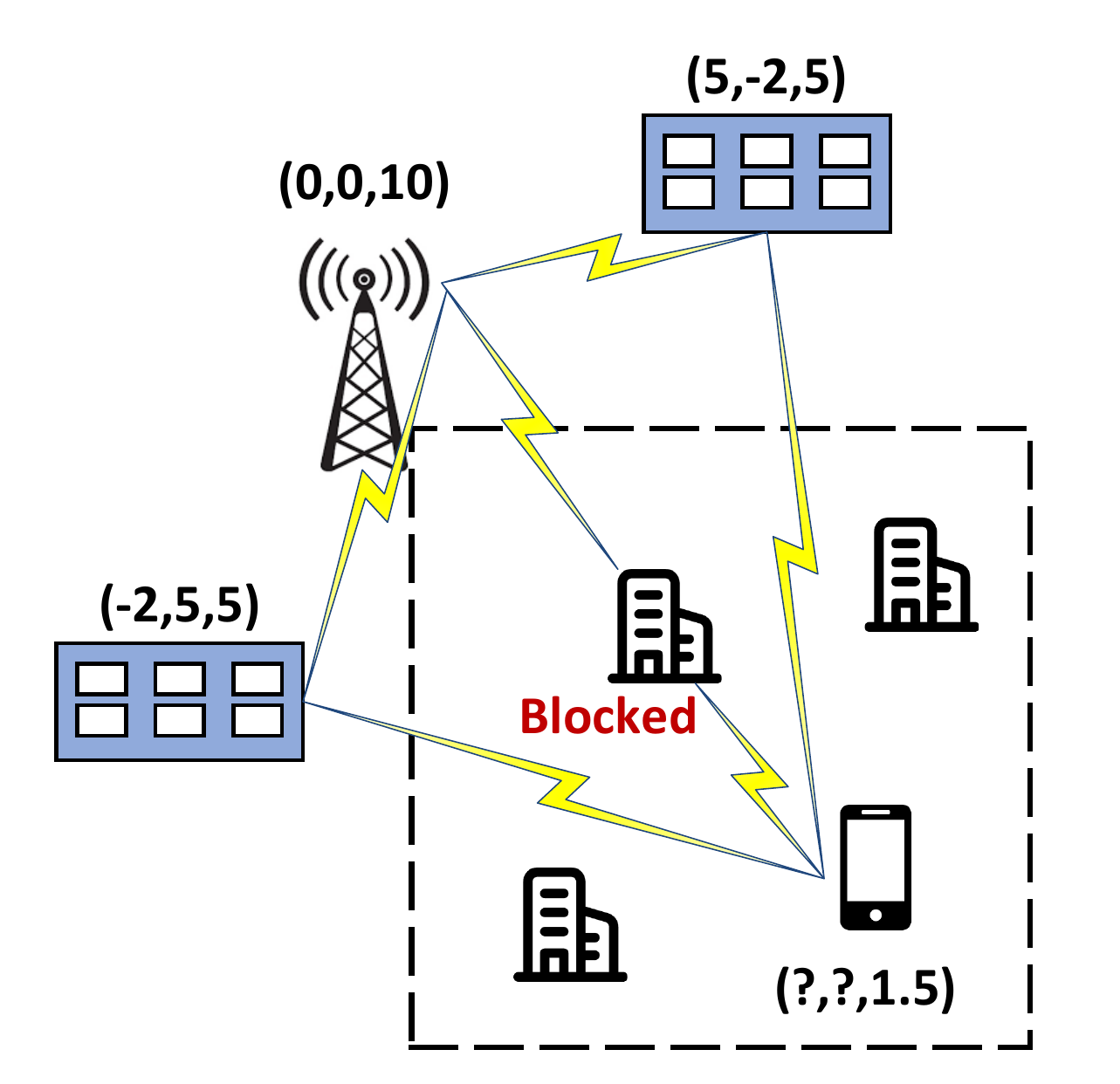}}\\
		\subfigure[]{\includegraphics[width=0.9\linewidth]{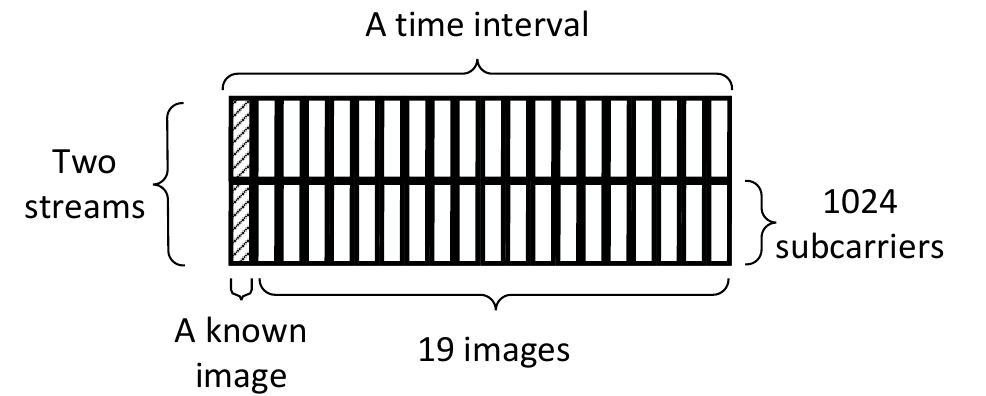}}
	\caption{(a) Channel scenario. (b) OFDM settings for RIS-SC.}
	\label{frame}
\end{figure}

In the given scenarios, the Rician channel model is employed with a delay spread of $L=2$. The channel consists of both line-of-sight (LoS) and non-line-of-sight (NLoS) components. The LoS component is position-dependent while the NLoS component follows a complex Gaussian distribution due to the multipath effect. The power ratio between the LoS and NLoS components is set to 10.
As shown in Fig. \ref{frame}(a), the user moves randomly within a $10\times 10$ area, while the BS is located at coordinates (0, 0, 10). Two RISs, namely RIS 1 and RIS 2, are positioned at coordinates (5, -2, 5) and (-2, 5, 5) respectively. The user's position remains constant during a given time interval.
It is important to note that the direct link between the BS and UT may be blocked as the user moves, resulting in the \textbf{blocked channel}. Conversely, the \textbf{ideal channel} refers to the scenario where a clear BS-UT path exists without any blockages.

The OFDM settings are illustrated in Fig. \ref{frame}(b). Each time interval transmits 20 images using a 2 × 2 MIMO system. Each image can utilize 2048 subcarriers. Consequently, each OFDM symbol consists of 1024 subcarriers, and two streams are served simultaneously.  Regarding encoding, an image is encoded into a total of 4096 bits, which can be arranged in either one stream using 16-QAM modulation or divided into two streams using 4-QAM modulation. During the initial 64 time intervals, 20 known images are transmitted to the receiver. Subsequently, in each following time interval, one known image is transmitted first while the remaining 19 images are received and evaluated for testing purposes. The SNR is set to 3 dB. The RL agents are trained using a batch size of eight. The training data is collected over multiple time intervals, and a training epoch is executed once eight known images are transmitted, with rewards calculated accordingly.

The OxfordPets dataset \cite{parkhi2012cats} is used, with 6,390 images in the training set and 1,000 images in the testing set. Additionally, the Camvid dataset \cite{brostow2009semantic} is utilized as an application scenario for the proposed method, consisting of 500 images in the training set and 201 images in the testing set. 
Among the competing methods, one is the conventional RL method, referred to as RL-Rate, which aims to enhance the sum-rate by controlling phase shifts. However, all the source-channel coding methods utilized in this comparison are based on semantic coding. This is because conventional coding methods like LDPC do not perform well under low SNR conditions and also require a higher number of bits. Semantic coding, on the other hand, is better suited for preserving the integrity of different semantic parts in the transmission. Despite this advantage, the RL-Rate method does not effectively exploit the different semantic parts since it treats them equally when optimizing the sum-rate.

  	\begin{table}[!h]
	\centering	
	\scriptsize
	\caption{The ACC and MSE performances of the SVC and conventional methods. All the agents are fully trained under two RISs and one UT. }

	\begin{tabular}{c|llll}   
		\toprule
			& RIS-SC&RIS-SC &RIS-SC&JPEG+LDPC  \\
   & (RL-ACC)&(RL-Rate) &(MAB)&(MAB)  \\\hline
ACC(10 dB SNR)&0.868 &0.868& 0.820&\textbf{0.953}\\ \hline
ACC(3 dB SNR) &\textbf{0.849}& 0.823&0.782&0.688 \\ \hline
MSE(10 dB SNR)&0.00671& 0.00667 &0.00832&\textbf{0}\\ \hline
MSE(3 dB SNR) &0.0217&\textbf{0.00987}&	0.0158 &0.284\\ 
		
		\bottomrule
	\end{tabular}
	\label{Metric1}
\end{table}

 Table \ref{Metric1} compares the performance of RIS-SC based on RL-Rate, RL-ACC, multi-armed bandit (MAB) \cite{9794416}, and other conventional methods such as JPEG+LDPC coding. The comparison considers an ideal channel condition with SNR set to 10 dB and 3 dB. All the RL-based methods demonstrate good performance and outperform MAB-based methods when the SNR is 10 dB. JPEG+LDPC employs a 1/2 code rate and consumes approximately 16 times the number of bits compared to the semantic coding method. While it can guarantee no bit errors at high SNR, JPEG+LDPC cannot effectively repair bit errors at an SNR of 3 dB, leading to a significant decrease in performance. The following subsections focus on the challenges posed by blocked channels and highlight the remarkable superiority of RIS-enhanced methods.

\subsection{Performances  under different channel conditions}

In this subsection, we discuss the requirements of RIS-enhanced methods and simulate scenarios with different numbers of controllable RISs and varying channel conditions. The blocked channel scenario is particularly interesting as it typically results in subchannels due to the channel matrix not being full-rank. Conventional methods transmit all the information through the good subchannels, whereas the proposed RIS-enhanced methods are more adaptive based on the user's requirements. Additionally, the RIS can directly improve the channel quality and enhance the task performance.

\begin{figure}[h]
	\centering  
	\subfigure[]{
		\label{Diff_RIS_ACC}
		{\includegraphics[width=0.9\linewidth]{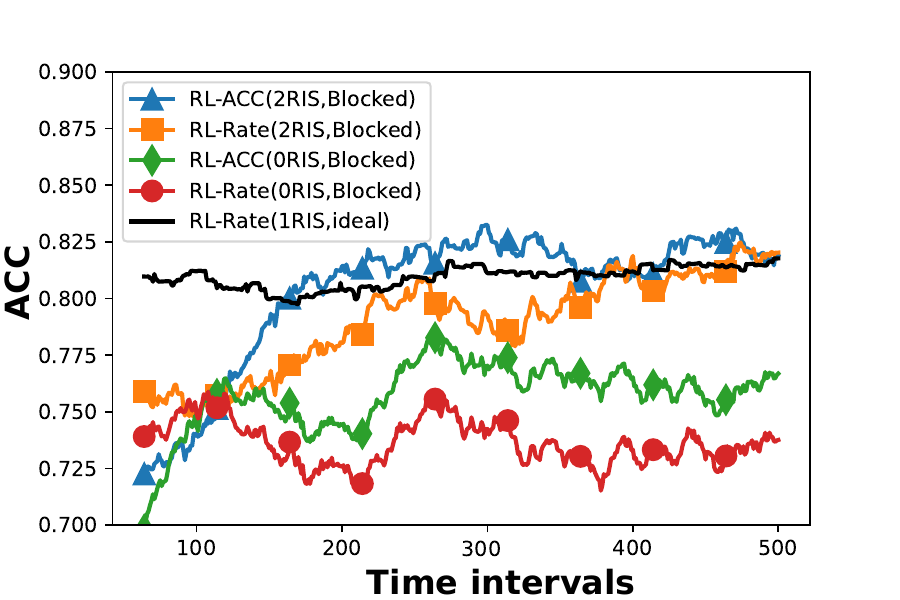}}}\\
\subfigure[]{
		\label{Diff_RIS_MSE}
		{\includegraphics[width=0.9\linewidth]{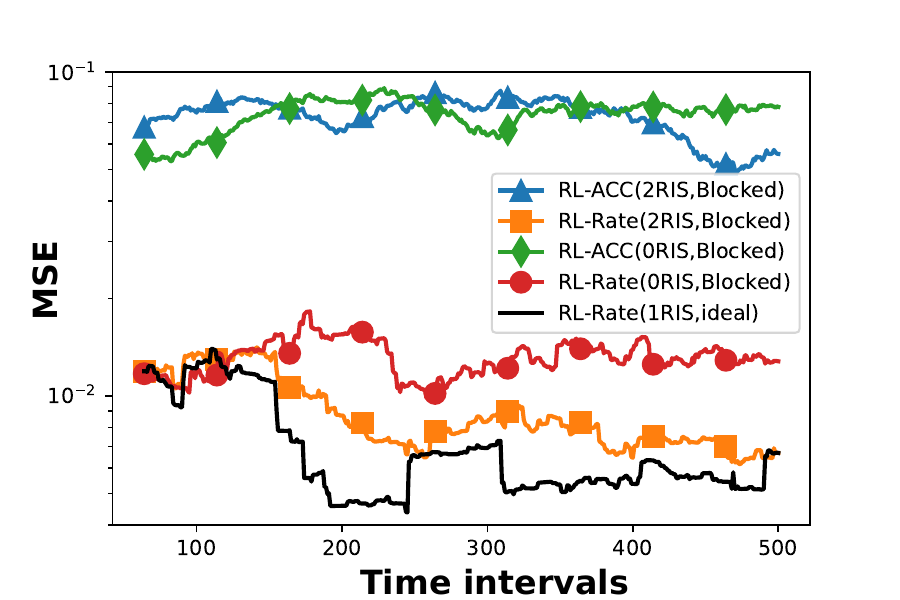}}}  
	\caption{Performances of the competing methods with different RISs and rewards, where BS-UT link is blocked. (a) Object classification accuracy performance.  (b) Global MSE performance.}
	\label{Diff_RIS}
\end{figure}

Fig. \ref{Diff_RIS} presents the experimental results under ideal and blocked channel conditions. In Fig. \ref{Diff_RIS_ACC}, RL-Rate (1 RIS, ideal) consistently achieves good classification performance. This is because the power of the direct BS-UT link and one BS-RIS-UT link is higher than that of two BS-RIS-UT links. RL-ACC (2 RIS, blocked)  initially performs worse than RL-Rate(1 RIS, blocked) due to the power difference, but later surpasses it. The adjustment of the RIS demonstrates its ability to improve the task performance. Despite no RIS adaption, RL-ACC (0 RIS, blocked) still performs better than RL-Rate (0 RIS, blocked) by properly arranging important semantic parts. However, RL-ACC achieves the task performance improvement by sacrificing the transmission of unimportant semantic parts, resulting in poorer MSE performance, as shown in Fig. \ref{Diff_RIS_MSE}. On the other hand, RL-Rate consistently achieves good MSE performance. The addition of RIS further enhances the transmission performance.

In conclusion, the proposed RIS-SC effectively allocates transmission resources based on the channel conditions. RL-Rate and RL-ACC perform similarly under ideal channel conditions, but RL-ACC achieves better classification performance under blocked channels. Furthermore, the addition of RIS provides an excellent opportunity to improve the task performance without the need to retrain the semantic transmission networks.

\subsection{Performance of RIS allocation under varying user requirements}
In this subsection, the BS only obtains the current chosen RIS rows of single users and aims to save the RIS elements. Varying channel conditions, where the ideal channel and UT-BS path have a 50\% probability of being blocked, will impact the results.

\begin{figure}[h]
	\centering  
	\subfigure[]{
		\label{RISeff_ACC}
		{\includegraphics[width=0.9\linewidth]{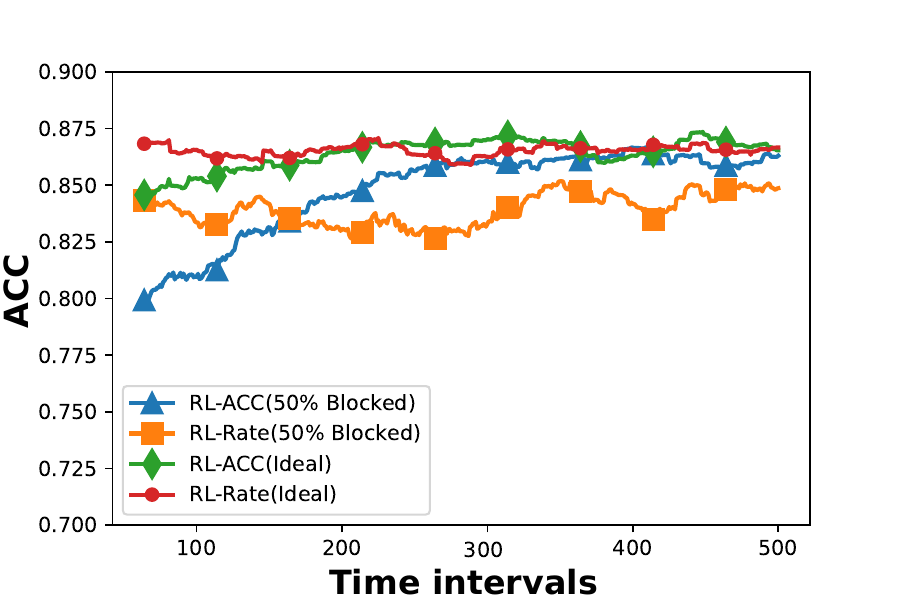}}}\\
\subfigure[]{
		\label{RISeff_MSE}
		{\includegraphics[width=0.9\linewidth]{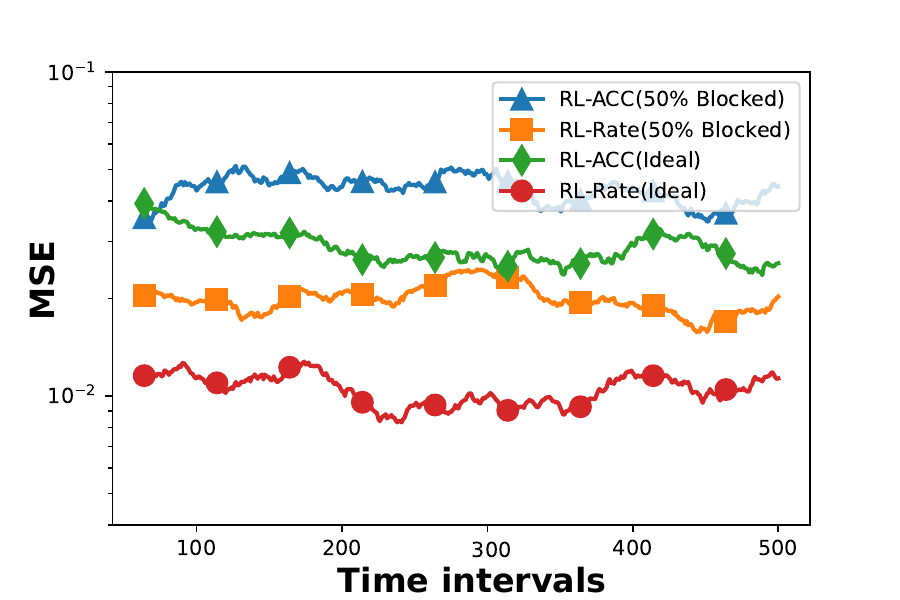}}} \\ 
  \subfigure[]{
		\label{RISeff}
		{\includegraphics[width=0.9\linewidth]{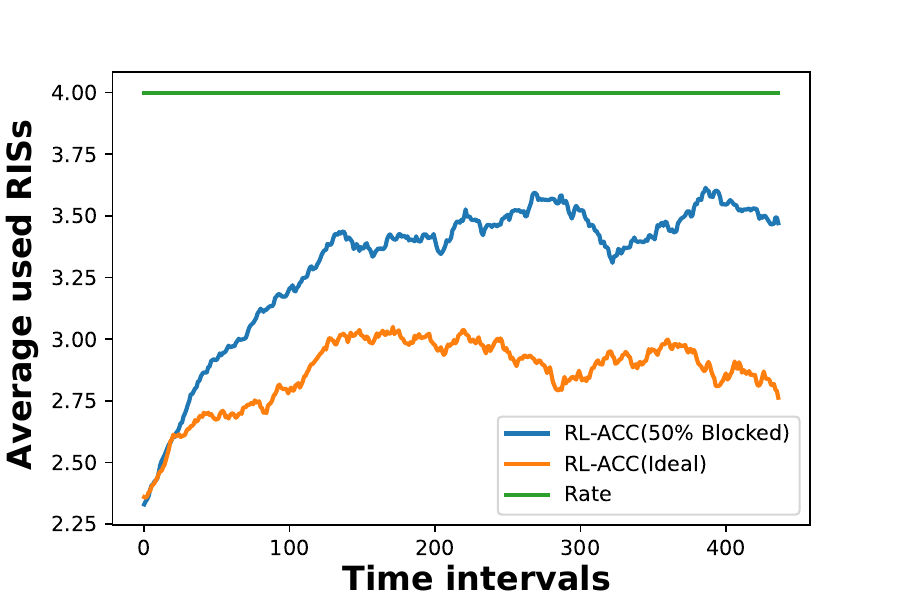}}}  
	\caption{The RL methods to save the RIS resources under different channel conditions: (a) ACC performance, (b) MSE performance, and (c) Average used number of the RIS rows.}
	\label{RISeff1}
\end{figure}

Fig. \ref{RISeff_ACC} shows that the ACC performances under 50\% blocked channels are worse than those under ideal channels. However, RL-ACC under 50\% blocked channels still manages to achieve results close to those under ideal channels through interaction with the environment. RL-Rate under ideal channels initially reaches good performance because its training process easily adjusts RIS elements for a high transmission rate. Subsequently, RL-ACC under ideal channels slightly surpasses RL-Rate. This phenomenon is more pronounced when the channel is blocked, with RL-ACC significantly outperforming RL-Rate under 50\% blocked channels. However, the MSE performances of RL-ACC are consistently worse than those of RL-Rate, as depicted in Fig. \ref{RISeff_MSE}. Specifically, RL-ACC under ideal channels exhibits a larger performance gap with RL-Rate compared to under 50\% blocked channels because the user requirement for ACC performance is easily met under ideal channels, and RL-ACC saves RIS resources instead of improving MSE performance.

\begin{figure}[h]
	\centering  

\subfigure[]{
		\includegraphics[width=0.3\linewidth]{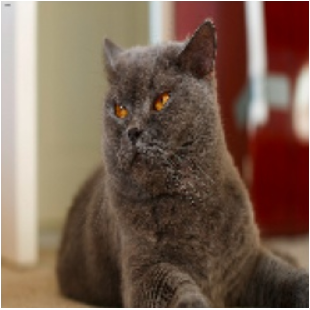}}
  \subfigure[]{
		\includegraphics[width=0.3\linewidth]{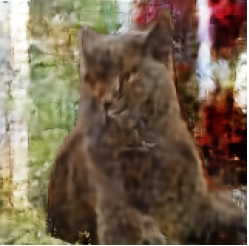}}
  \subfigure[]{
  \includegraphics[width=0.3\linewidth]{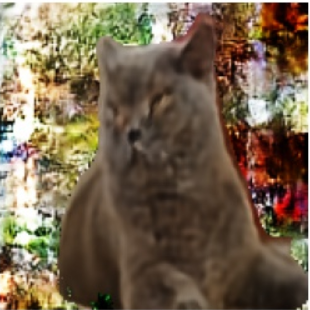}}

  \subfigure[]{
		\includegraphics[width=0.3\linewidth]{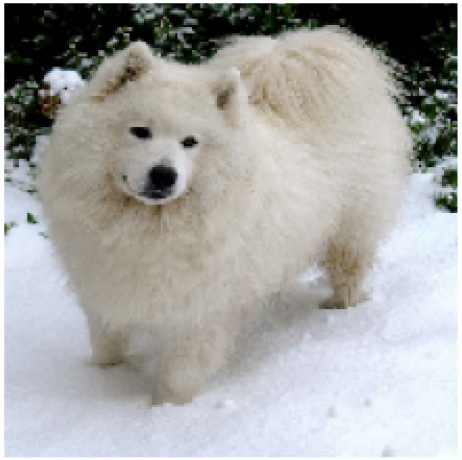}}
  \subfigure[]{
		\includegraphics[width=0.3\linewidth]{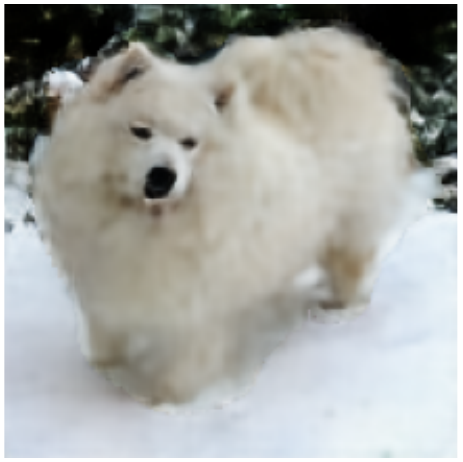}}
  \subfigure[]{
  \includegraphics[width=0.3\linewidth]{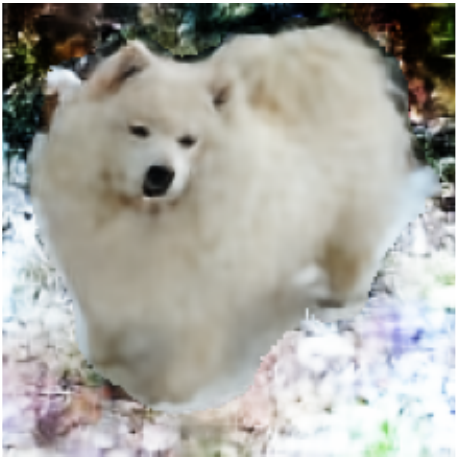}}
 
	\caption{The examples of the proposed method. (a)(b)(c) are the transmitted, RL-Rate received, RL-ACC received images using all the RISs under  blocked channels, respectively. (d)(e)(f) are the transmitted, RL-Rate received, RL-ACC received images saving RIS resources under ideal channels, respectively.}
	\label{Diff_ex}
\end{figure}

Fig. \ref{RISeff} demonstrates the effectiveness of the proposed method in terms of RIS row usage. RL-ACC under 50\% blocked channels requires an average of 3.5 RIS rows, which is slightly more than the average of 3.0 RIS rows under ideal channels. RL-ACC can achieve the task performance with fewer RIS rows compared to RL-Rate, which consistently utilizes all RIS rows to achieve a high transmission rate. Furthermore, RL-ACC can learn to use the limited RIS resources to achieve good task performance.

Lastly, to illustrate the impact of changing requirements, image examples are provided in Fig. \ref{Diff_ex}. Figures \ref{Diff_ex}(b) and \ref{Diff_ex}(c) depict the performance of RL-ACC and RL-Rate with two RISs under blocked channels. RL-Rate assumes equal transmission and restoration of both objects and background. RL-ACC achieves better object performance while meeting the background performance requirements. In contrast, Figures \ref{Diff_ex}(e) and \ref{Diff_ex}(f) are received under ideal channel conditions, and RL-Rate reaches good overall performance. When all RISs are used, RL-ACC can achieve a similar performance to RL-Rate. However, RL-ACC prioritizes satisfying the background performance to save RIS resources.

The above experiments highlight the superiority of the proposed method in improving task performance and resource utilization. However, the transmission quality of other semantic parts may be sacrificed, leading to degraded performance when user requirements change. In the following subsection, we consider changing requirements of different users.

\subsection{Performances under varying user requirements}

In this subsection, two users are served by the same BS and their requirements are taken into consideration when controlling the RIS elements. Furthermore, the users' requirements change over time. User 1's requirement is ACC performance during time intervals 0-500 and 1500-2000, while it switches to MSE performance during time intervals 500-1500. User 2's requirement is MSE performance during time intervals 0-1000, which changes to ACC performance during time intervals 1000-2000. As a result, the two users may have the same or different requirements, and the proposed method should be able to adapt to this.

\begin{figure}[t]
	\centering  
	\subfigure[]{
		\label{Diff_Re_ACC}
		{\includegraphics[width=0.9\linewidth]{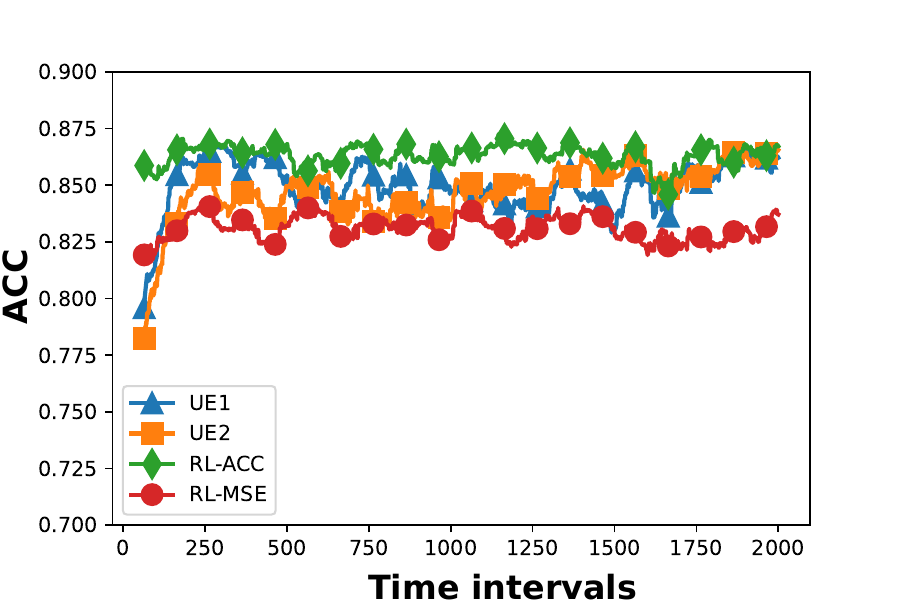}}}\\
\subfigure[]{
		\label{Diff_Re_MSE}
		{\includegraphics[width=0.9\linewidth]{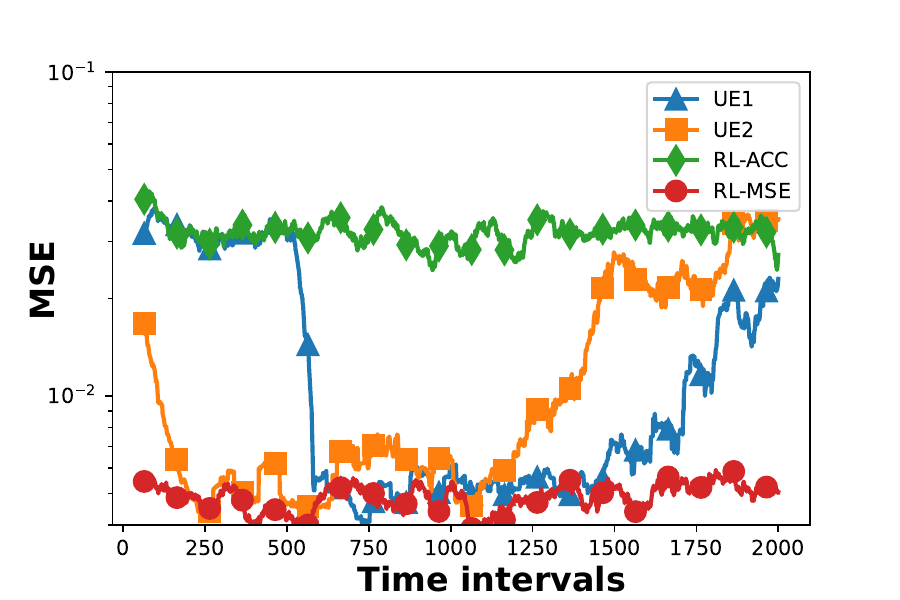}}}
	\caption{Performances under changing user requirements: (a) ACC performance, and (b) MSE performance}
	\label{Diff_Re}
\end{figure}

 Fig. \ref{Diff_Re} shows the corresponding results. RL-ACC and RL-MSE are fully trained offline, and their rewards remain unchanged during testing and online adaptation. Therefore, RL-ACC and RL-MSE represent the best ACC and MSE performances under those respective environments. User 1 achieves the best ACC performance during time intervals 0-500 but sacrifices MSE performance to save resources for User 2, who is adapted to reach RL-MSE in terms of MSE performance. Then, User 1's requirement changes, and its MSE performance rapidly improves due to online adaptation. As a result, both users have similar MSE performances during time intervals 500-1000, but their ACC performances are not optimal. During time intervals 1000-1500, User 2 requires object recognition task performance, leading to a decline in its MSE performance to save resources, while its ACC performance surpasses that of User 1. Finally, both users achieve the best ACC performance according to their respective requirements by sacrificing MSE performance.

 The experiments demonstrate that the proposed method can balance different task performances guided by changing rewards and can quickly adapt through online adjustments. The method aims to meet all user requirements under limited bandwidth, which may impact unimportant semantic parts. However, these unimportant semantic parts may be visually unacceptable, and this issue will be discussed in the following subsection.

\subsection{Reconstruction for blocked semantic parts}
In this subsection, the proposed methods are tested using pictures of a street, providing new application scenarios. Different users have different requirements for objects in the street, and two classic applications of extended reality are considered. For remotely operated vehicles, high-quality transmission of cars, signs, and roads is important, while online tourists may be more interested in buildings, plants, and the sky. Therefore, the rewards used are MSE performances of the required parts and the global images. Additionally, the effectiveness of the proposed image reconstruction method is demonstrated. The simulation processes follow the same procedures as the previous experiments, with ACC and MSE rewards replaced by MSE (required) and MSE (global) rewards.
 
\begin{figure}[t]
	\centering

		{\includegraphics[width=0.9\linewidth]{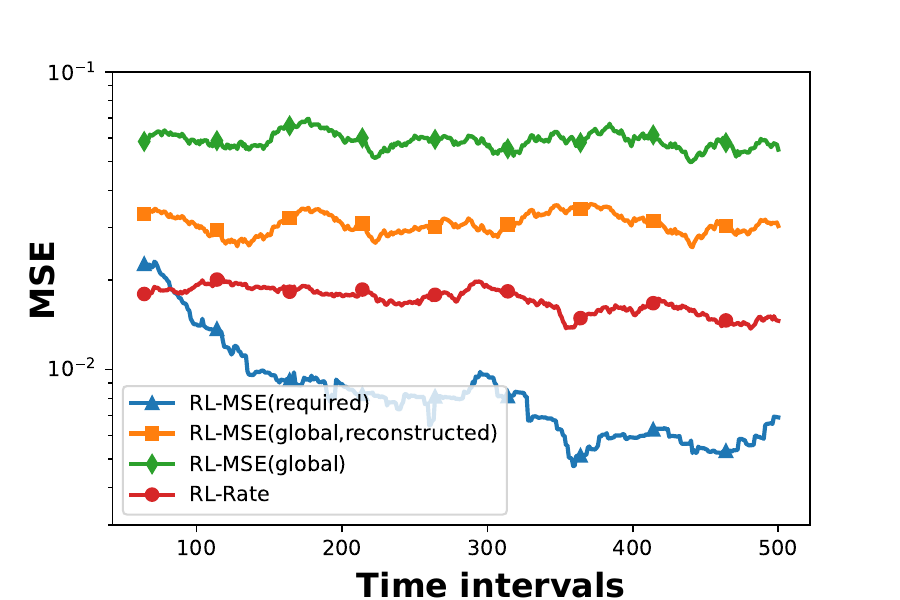}}

	\caption{MSE performances of the required parts for online tourist and the global image, as well as the reconstructed image, are tested. }
	\label{Rec}
\end{figure}

 As online tourists, the crowd in the street may be unimportant because the focus is on classic architecture and natural beauty. The simulation results in Fig. 11 show that RL-MSE (required) consistently achieves better performance than RL-MSE (global). This indicates that the required semantic parts are well transmitted through the RIS-enhanced systems. In contrast, RL-Rate achieves MSE performance between RL-MSE (required) and RL-MSE (global) because traditional methods cannot address user requirements effectively. Moreover, the blocked semantic parts can be reconstructed, and the MSE performance of RL-MSE (global, reconstructed) is better than RL-MSE (global), although still worse than RL-Rate. This demonstrates the effectiveness of the reconstruction method, but the reconstructed part differs from the original one. Therefore, a thorough evaluation of the proposed reconstruction method cannot be solely based on MSE performance.

\begin{figure}[t]
	\centering  
	\subfigure[]{
		
		{\includegraphics[width=0.3\linewidth]{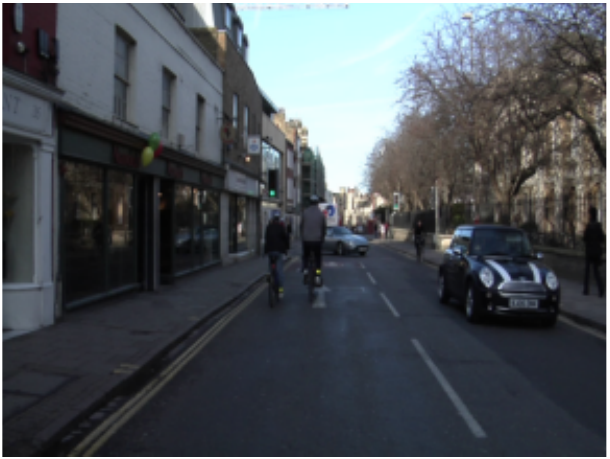}}}
\subfigure[]{
	
		{\includegraphics[width=0.3\linewidth]{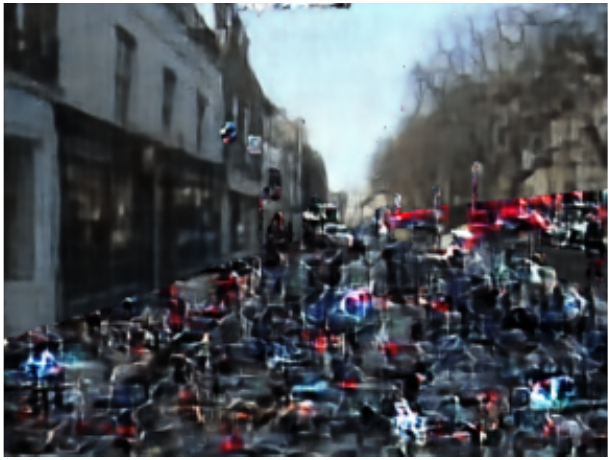}}}
  \subfigure[]{
		
		{\includegraphics[width=0.3\linewidth]{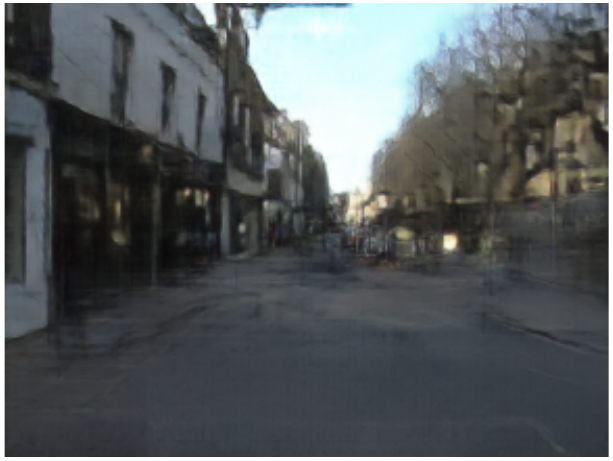}}}
  \subfigure[]{
		
		{\includegraphics[width=0.3\linewidth]{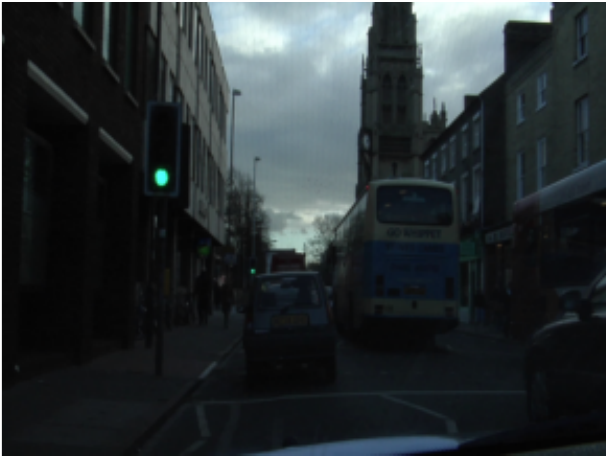}}}
  \subfigure[]{
		
		{\includegraphics[width=0.3\linewidth]{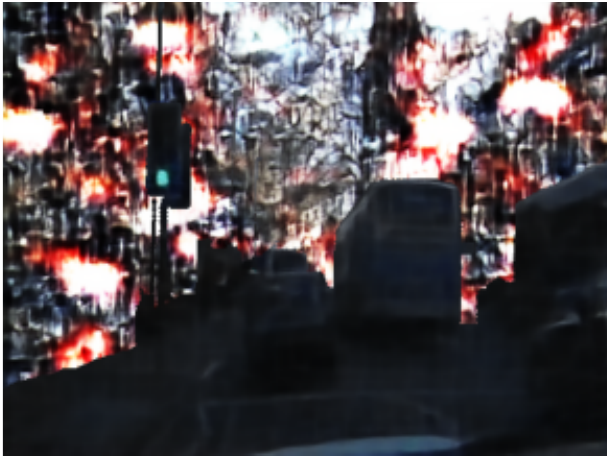}}}
  \subfigure[]{
		
		{\includegraphics[width=0.3\linewidth]{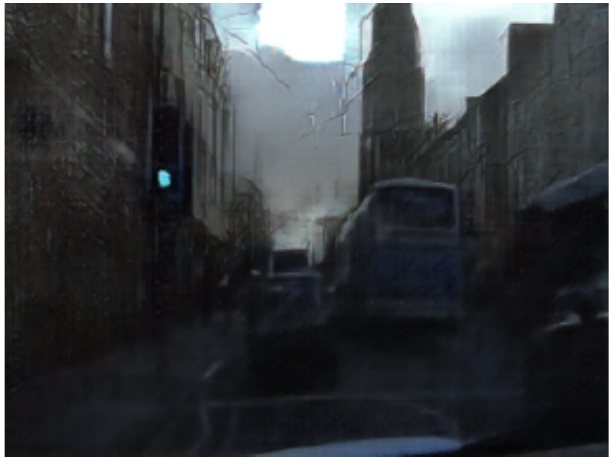}}}
	\caption{Examples of reconstructed images: (a), (b), and (c) are the transmitted, received, and reconstructed images for tourists. (d), (e), and (f) are the transmitted, received, and reconstructed images for remote drivers.}
	\label{Rec_example}
\end{figure}

 To explain the reconstructed images, examples are shown in Fig. \ref{Rec_example}. Consider an extremely poor environment where all transmission resources are utilized to transmit required parts. In this case, the generator must guess the blocked part based on the required part. For online tourists, the cars and bicycles in the road are absent because the generator cannot reconstruct these details from noise  (see Fig. \ref{Rec_example}(c)). However, these missing details are acceptable for online tourists since they focus primarily on the overall view of the city. For remotely operated vehicles, the buildings, sky, and trees may be artificially generated, but these parts have no impact on remote driving (see Fig. \ref{Rec_example}(f)).

Overall, denoising or generating images has been extensively studied, and these technologies have the potential to enhance unimportant semantic parts without occupying additional transmission resources. However, denoising or generating an image requires the neural network to guess information from surrounding pixels or other good parts, reducing reliability and not being suitable for important semantic parts. Important parts should be transmitted and protected effectively, and RIS-enhanced methods offer a promising opportunity to adjust channel environments accordingly.

\section{Conclusion}
\label{s6}

This study introduces a novel framework called RIS-SC, which aims to enhance semantic communication in MIMO systems, particularly in challenging channel conditions. The framework utilizes RL techniques to dynamically adjust the phase shifts of reconfigurable intelligent surfaces (RISs) and optimize transmit bit sequences based on user requirements.  Because the proposed method exhibits remarkable adaptability without retraining semantic communication networks, it can be easily applied in different semantic methods. 
Next, this paper investigates the utilization of RIS, where the RIS resources are effectively allocated among users with dynamic requirements. User requirements are fed back online, allowing the proposed method to quickly learn and satisfy the users' needs within the constraints of limited RIS resources. 
Finally, it is observed that under poor channel conditions, certain semantic parts may not be accurately received, potentially degrading the visual quality of the received images. To address this issue, a reconstruction method  is proposed to ensure the preservation of visual quality. Overall, the proposed RIS-SC is a highly mobile and cost-effective method in the physical layer to safeguard user requirements under extreme channel conditions. It also offers a good service for meeting different users' preferences and expectations simultaneously.

	\bibliographystyle{IEEEtran}
	\bibliography{bibtex0320}
	
	%
	
	
	
\end{document}